# A study of co-movements between USA and Latin American stock markets: a cross-bicorrelations perspective

Semei Coronado[a], Omar Rojas[b], Rafael Romero-Meza[c] & Francisco Venegas-Martínez[d]

[a] *Department of Quantitative Methods, Universidad de Guadalajara, Zapopan, México,* semeic@cucea.udg.mx
[b] *School of Business and Economics, Universidad Panamericana, Guadalajara, México.* orojas@up.edu.mx
[c] *PKF Chile Finanzas Corporativas, Región Metropolitana, Chile,* rromero@pfkchile.cl
[d] *Superior School of Economics, Instituto Politécnico Nacional, Ciudad de México, México, e-mail:* fvenegas1111@yahoo.com.mx

**Abstract**
In this paper we use the Brooks and Hinich cross-bicorrelation test in order to uncover nonlinear dependence periods between USA's Standard and Poor's 500 (SP500), used as benchmark, and six Latin American stock markets indexes: Mexico (BMV), Brazil (BOVESPA), Chile (IPSA), Colombia (COLCAP), Peru (IGBVL) and Argentina (MERVAL). We have found windows of nonlinear dependence and co-movement between the SP500 and the Latin American stock markets, some of which coincide with periods of crisis, giving way to a possible contagion or interdependence interpretation.

## 1 Introduction

The study of the transmission of shocks from one country to others and the cross-countries correlations and co-movements, beyond any fundamental link, has long been a subject of interest to researchers in economics and finance, as well as to fund managers and traders since it has, amongst others, important implications for asset pricing and portfolio allocations.

Recently, co-movements between financial markets have been studied, with emphasis on stock market indexes' returns, through econometric or time series models that allow for a better understanding of the behavior of markets, especially during times of crisis. A lot of this research has studied interdependencies and co-movements from the point of view of contagion, cf. [1-6]. There are many approaches to analyze the co-movements between financial markets, some of which use quantitative tools borrowed mostly from physics and computational sciences, cf. [7-9].

According to [2], if two markets are highly correlated, and the correlation does not increase in one of the markets after a financial crisis, but on the contrary, there is a continuous variation in its co-movement, then both markets are highly interdependent and contagion cannot be considered. Thus, for these authors the privileged dimension is a linear dependency. However, there is an important line of research that emphasizes the need for an empirical verification of nonlinear univariate and multivariate dependencies. There are at least two reasons for the popularity of this line of research, according to [10]. First, if evidence of nonlinearity is found in the residuals of a linear model, this must cast doubt on the adequacy of the linear model as an adequate representation of the data. Second, if the nonlinearity is present in the conditional first moment, it may be possible to devise a trading strategy based on nonlinear models, which is able to yield higher returns than a buy-and-hold rule.

In order to explore this –less known and less obvious– nonlinear relationships between financial markets and how they co-move, in this work we use the Brooks and Hinich [11] nonlinearity test, which allows for cross-correlations and cross-bicorrelations between pairs of series, using bivariate autoregressive vectors in high frequency. These tests can be viewed as natural multivariate extensions of Hinich's portmanteau bicorrelation and whiteness statistics [12], which search for nonlinear co-features between time series. The advantage in using the cross-bicorrelation test is that it addresses the specific window frames in which the nonlinear dependence is present and also signals the direction of the nonlinear dependence, which is not provided by the Granger causality test

Both univariate [13-19] and multivariate [10; 20-22] tests has been successfully applied to analyze the nonlinear behavior of different financial and economic time series. However, to the best of our knowledge, this is the first time that such multivariate nonlinear test is used to uncover how stock markets co-move.

In this paper, we use the Brooks and Hinich nonlinearity test in order to uncover the cross-covariances and cross-correlations between USA's Standard and Poor's 500 (SP500), used as benchmark, and six Latin American stock markets indexes: Mexico (BMV), Brazil (BOVESPA), Chile (IPSA), Colombia (COLCAP), Peru (IGBVL) and Argentina (MERVAL). We have found windows of nonlinear dependence between the SP500 and the Latin American stock markets, some of which coincide with periods of crisis, giving way to a possible contagion or interdependence interpretation.

The plan of the paper is as follows: Section 2 presents the data sample. Section 3 describes the methodology used. Section 4 reports the empirical results. Finally, concluding remarks are presented in Section 5.

## 2 The data

For this study we consider daily returns of seven stock market indexes, namely the Standard and Poor 500 (SP500) from USA is taken as a baseline market for comparison against six Latin American stock market indexes: Mexico (BMV), Brazil (BOVESPA), Chile (IPSA), Colombia (COLCAP), Peru (IGBVL) and Argentina (MERVAL). Daily closing prices from January 2$^{nd}$, 2003 to January 8$^{th}$, 2015, for a total of 3025 observations of each index were obtained from *Bloomberg*. The data was sampled for this period of time in order to capture the effects that the USA might have had on the Latin American stock markets during the sub-prime financial crisis and to have a broad view of other possible cross-bicorrelation phenomena. Prices where transformed into series of continuously compounded percentage returns, i.e. $r_t = 100\,(\ln(p_t) - \ln(p_{t-1}))$, where $p_t$ is the daily closing price of the stock market on day $t$. Table 1 presents summary statistics for these returns. The statistics are consistent, as expected, with some of the stylized facts of financial time series [23-24]. In particular, the kurtosis indicates that return distributions are leptokurtic. Furthermore, the Jarque-Bera statistic (JB) confirms returns not normally distributed. Although the results of the KPSS test for seasonality are not listed, it does not reject the null hypothesis of seasonality or the results of the ADF and PP tests, under the null hypothesis of a unit root. Both tests with a 5% significance level are available upon request.

Table 1.
Summary statistics for the returns

|  | Mean | Min | Max | Std-Dev | Skewness | Kurtosis | JB |
|---|---|---|---|---|---|---|---|
| SP500 | 0.03 | -9.47 | 10.96 | 1.23 | -0.32 | 14.15 | 15727 |
| BMV | 0.06 | -7.27 | 10.44 | 1.27 | 0.06 | 9.24 | 4905 |
| BOVESPA | 0.06 | -12.10 | 13.68 | 1.79 | -0.07 | 7.94 | 3073 |
| IPSA | 0.05 | -7.17 | 11.80 | 1.03 | -0.03 | 13.19 | 13100 |
| COLCAP | 0.08 | -13.25 | 18.13 | 1.36 | -0.12 | 23.85 | 54822 |
| IGBVL | 0.07 | -13.29 | 12.82 | 1.54 | -0.52 | 13.18 | 13191 |
| MERVAL | 0.10 | -12.95 | 10.43 | 1.99 | -0.57 | 7.03 | 2216 |

Source: The authors using data from *Bloomberg*

## 3 Methodology

The Brooks and Hinich cross-bicorrelation test allows us to detect any presence of nonlinear dependence between two time series. The data sample is of size $N$, with two stationary time series $x(t_k)$ and $y(t_k)$. As we are working with the first percentual logged differences and small sub-samples of the total series, to suppose stationarity is more than reasonable. Each series is separated into equal length non-overlapping moving time windows or frames, where $t$ is an integer and $k$ represents the $k$-th window and, both series are jointly covariance stationary, which have been standardized. The test's null hypotheses states that the two series $x(t_k)$ and $y(t_k)$ are independent and pure white noise, against an alternative hypothesis that states that the series have cross-covariances, $C_{xy}(r) = E[x(t_k)y(t_k + r)]$, or any of the cross-bicovariances, $C_{xxy}(r,s) = E[x(t_k)x(t_k + r)y(t_k + s)]$, different from zero.

Under the null hypothesis, $C_{xy}(r) = 0$ y $C_{xxy}(r,s) = 0$ for every $r, s$ except when $r = s = 0$. It is well known that a pair of series are dependent if there is a second or third order lagged dependence between the two series, then $C_{xy}(r) \neq 0$ or $C_{xxy}(r,s) \neq 0$ for at least one value of $r$ or a pair of values of $r$ and $s$, respectively. The following statistics give the $r$ sample $xy$ cross-correlation and the $r, s$ sample $xxy$ cross-bicorrelation, respectively

$$C_{xy}(r) = (N-r)^{-1} \sum_{t=1}^{N-r} x(t_k)y(t_k + r), \qquad r \neq 0$$

and

$$C_{xxy}(r,s) = (N-m)^{-1} \sum_{t=1}^{N-m} x(t_k)x(t_k + r)y(t_k + s)$$

where $m = \max(r,s)$.

The cross-bicorrelations can be seen as a correlation between the current value of one of the series and the value of the previous cross-correlation between the two series. The second-order test does not include contemporaneous terms, and is conducted on the residuals of an $AR(2)$ fit to filter out the univariate autocorrelation structure so that contemporaneous correlations will not cause rejections. For the third-order test, we estimate the test on the residuals of a $VAR(2)$ model containing a contemporaneous term in one of the equations (the order $p$ of the $AR(p)$ and $VAR(p)$ models is chosen optimally using the Schwartz (BIC) criterion). The motivation for this pre-whitening step is to remove any traces of linear correlation or cross-correlation so that any remaining dependence between the series must be of a nonlinear form. Let $L = N^c$ where $0 < c < 0.5$ (for our case of study we use $c = 0.4$, $N = 3025$, and thus we have 121 non-overlapped windows of length 25 days). The test statistics for non-zero cross-correlations and cross-bicorrelations are

$$H_{xy}(N) = \sum_{r=0}^{L}(N-r)C_{xy}^2(r)$$

and

$$H_{xxy}(N) = \sum_{s=-L}^{L}\sum_{r=1}^{L}(N-m)C_{xxy}^2(r,s), \quad -s \neq -1,1,0$$

respectively. These tests are joint or composite tests for cross-correlations and cross-bicorrelations, where the number of proved correlations is $L$ and the number of cross-bicorrelations tested for is $L(2L-1)$. According to [12], $H_{xy}$ and $H_{xxy}$ are asymptotically $\chi^2$ with $L$ and $L(2L-1)$ degrees of freedom, respectively, as $N \to \infty$.

## 4 Empirical results

In Table 2, we report the results for the cross-bicorrelation test. All tests are run taking SP500 as the benchmark for comparison, since the effects of USA on Latin America are the ones we wanted to test. We present the number and percentage of significant (at the 5% level) cross-bicorrelation windows, correlation for all windows and the correlation for the largest window. As can be seen, the countries with the most significant cross-bicorrelation windows are Brazil (BOVESPA), Argentina (MERVAL) and Peru (IGBVL), with 28.1%, 25.6% and 24.9% of significant windows, respectively. On the other hand, Colombia (COLCAP), Chile (IPSA) and Mexico (BMV) are the countries with less significant windows (14.0%, 14.9% and 16.5%, respectively). As for the correlations for all windows, most countries present a correlation between 0.70 and 0.79, with the exception of Argentina (0.55). Furthermore, the country with an episode of largest correlation (0.50) was Argentina, whereas the one with the lowest correlation for a single episode was Peru (-0.01). These results shed light on the degree of dependence and co-movement between economies.

Table 2.
Number and percentage of significant (at the 5% level) cross-bicorrelation windows and correlations between *xxy* and *yyx*, and values of most significant bicorrelations. All cross-correlations and correlations are against SP500.

| Series | Significant cross-bicorrelation windows | Correlations (*xxy*, *yyx*) for all windows | Correlation (*xxy*, *xy*) for largest window |
|---|---|---|---|
| BMV | 20 (16.5%) | 0.79 | 0.19 |
| BOVESPA | 34 (28.1%) | 0.70 | 0.12 |
| IPSA | 18 (14.9%) | 0.75 | 0.08 |
| COLCAP | 17 (14.0%) | 0.73 | 0.10 |
| IGBVL | 29 (24.0%) | 0.74 | -0.01 |
| MERVAL | 31 (25.6%) | 0.55 | 0.50 |

Source: The authors.

Table 3.
Dates of significant cross-correlation windows between SP500 and the corresponding Latin American stock market, labeled as follows: BMV (A), BOVESPA (B), IPSA (C), COLCAP (D), IGBVL (E) and MERVAL (F). All windows are of 25 labor days of length, for a total of 121 windows. The number in the A-F columns corresponds to the number of significant window.

| Year | Starting date | Finishing date | A | B | C | D | E | F |
|---|---|---|---|---|---|---|---|---|
| 2003 | 1/3 | 2/7 |  | 1 |  |  | 1 | 1 |
| 2003 | 2/10 | 3/17 |  | 2 |  |  |  | 2 |
| 2003 | 3/18 | 4/22 |  | 3 |  |  |  | 3 |
| 2003 | 5/29 | 7/2 |  |  |  |  |  | 5 |
| 2003 | 7/3 | 8/7 |  |  |  |  |  | 6 |
| 2003 | 9/15 | 10/17 |  |  |  |  |  | 8 |
| 2006 | 6/26 | 7/31 | 36 |  |  |  |  |  |
| 2007 | 1/31 | 3/7 |  |  |  |  |  |  |
| 2007 | 3/8 | 4/12 |  |  |  |  |  |  |
| 2007 | 5/18 | 6/22 |  |  |  |  | 45 |  |
| 2007 | 7/31 | 9/4 | 47 | 47 |  | 47 | 47 |  |
| 2007 | 10/10 | 11/13 |  | 49 |  |  |  | 49 |
| 2007 | 11/14 | 12/19 | 50 | 50 | 50 |  | 50 | 50 |
| 2008 | 12/20 | 1/28 | 51 | 51 | 51 |  | 51 |  |
| 2008 | 1/29 | 3/4 |  | 52 | 52 |  | 52 |  |
| 2008 | 3/5 | 4/9 | 53 | 53 | 53 | 53 | 53 | 53 |
| 2008 | 4/10 | 5/14 |  | 54 |  | 54 |  |  |
| 2008 | 5/15 | 6/19 |  | 55 |  |  |  | 55 |
| 2008 | 6/20 | 7/25 |  | 56 |  | 56 | 56 | 56 |
| 2008 | 7/28 | 8/29 |  | 57 |  |  | 57 |  |
| 2008 | 9/2 | 10/6 | 58 | 58 | 58 | 58 | 58 | 58 |
| 2008 | 10/7 | 11/10 | 59 | 59 | 59 | 59 | 59 | 59 |
| 2008 | 11/11 | 12/17 | 60 | 60 | 60 | 60 | 60 | 60 |

Source: The authors.

In Table 3 we present the dates of significant cross-bicorrelation windows between the SP500 benchmark and the six Latin American stock market indexes, labeled in the following way: BMV (A), BOVESPA (B), IPSA (C), COLCAP (D), IGBVL (E) and MERVAL (F). All windows are of 25 labor days of length, for a total of 121 windows. During 2003, Brazil and Argentina showed significant cross-correlation windows with USA. For the years of 2004 to April 2007, the markets do not co-move, with the exception of a significant window from 6/26 to 7/31 between SP500 and BMV. In the middle of 2007 the effects of USA are visible on some Latin American countries: Mexico, Brazil and Peru are affected earlier than Chile, Colombia and Argentina. However, the effects of the sub-prime financial crisis are felt on all countries from September 2008 lasting till July 209 (Table 3 (cont.)). In the years of 2009-2011, there is not much co-movement, Chile being the exception with significant cross-bicorrelation windows during these years. Another block of co-movement is visible from July to December 2011, which might have happened due to the European sovereign debt crisis and some of the concerns over USA's slow economic growth and its credit rating being downgraded. From 2012 to January 2015, the countries that continue to show significant cross-bicorrelations with USA are Brazil and Argentina.

Table 3 (cont.)

| Year | Starting date | Finishing date | A | B | C | D | E | F |
|---|---|---|---|---|---|---|---|---|
| 2009 | 12/17 | 1/23 | 61 | 61 | 61 | 61 | 61 | 61 |
| 2009 | 1/26 | 3/2 | 62 | 62 | 62 | 62 | 62 | 62 |
| 2009 | 3/3 | 4/6 | 63 | 63 | 63 | 63 | 63 | 63 |
| 2009 | 4/7 | 5/12 | 64 | 64 | 64 | 64 | 64 | 64 |
| 2009 | 5/13 | 6/17 | 65 | 65 | 65 | | 65 | 65 |
| 2009 | 6/18 | 7/24 | 66 | | 66 | 66 | 66 | 66 |
| 2009 | 7/24 | 8/27 | | 67 | | | | 67 |
| 2009 | 10/5 | 11/6 | 69 | 69 | | | | |
| 2010 | 1/22 | 2/26 | | | | | 72 | |
| 2010 | 4/6 | 5/10 | | | | | 74 | |
| 2010 | 5/11 | 6/15 | 75 | 75 | 75 | | 75 | 75 |
| 2010 | 6/16 | 7/21 | | 76 | | | | 76 |
| 2010 | 8/26 | 9/30 | | | | | 78 | |
| 2011 | 2/24 | 3/30 | | | | | 83 | |
| 2011 | 5/6 | 6/10 | | | | | 85 | |
| 2011 | 7/19 | 8/22 | 87 | 87 | 87 | 87 | 87 | 87 |
| 2011 | 8/23 | 9/27 | 88 | 88 | 88 | 88 | 88 | 88 |
| 2011 | 9/28 | 11/1 | 89 | 89 | 89 | 89 | 89 | 89 |
| 2011 | 11/2 | 12/7 | 90 | 90 | 90 | 90 | 90 | 90 |
| 2012 | 12/8 | 1/13 | | 91 | | | | 91 |
| 2012 | 1/17 | 2/21 | | 96 | | | | |
| 2013 | 3/28 | 5/2 | | | | | 104 | |
| 2013 | 6/10 | 7/15 | | 106 | | | | |
| 2014 | 1/13 | 2/18 | | 112 | | | | 112 |
| 2014 | 9/23 | 10/27 | | 119 | | | | 119 |
| 2015 | 12/3 | 1/8 | | | | 121 | | 121 |

In Fig. 1 and Fig. 2, we plot the $(1 - p)$-values of the significant cross-bicorrelation windows between SP500 and Mexico, Brazil and Colombia (Fig. 1), and Chile, Peru and Argentina (Fig. 2). These windows correspond to the ones reported in Table 3 and 4. It is clear how there are two main periods of nonlinear dependence between USA and Latin American stock markets: 2008-2009 and 2011. Brazil and Argentina show cross-bicorrelations with USA in 2003.

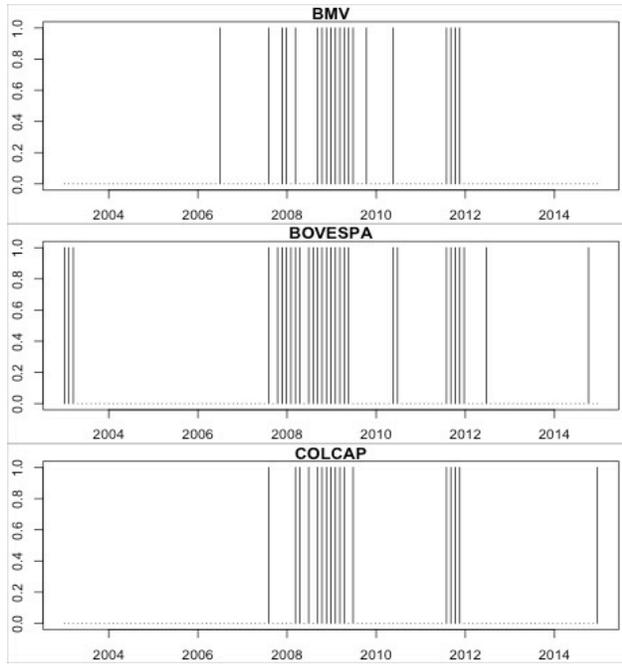

Figure 1. ($1-p$)-values of the significant cross-bicorrelation windows between SP500 and Mexico, Brazil and Colombia, respectively.
Source: The authors.

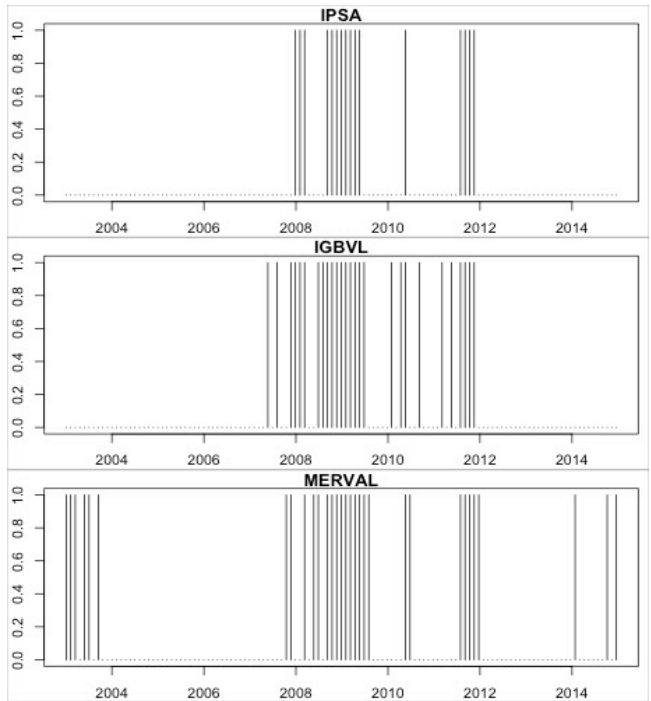

Figure 2. ($1-p$)-values of the significant cross-bicorrelation windows between SP500 and Chile, Peru and Argentina, respectively.
Source: The authors.

In Fig. 3 we plot the normalized prices for the SP500, BOVESPA and COLCAP indexes. The SP500 is the benchmark and it is compared with BOVESPA that presents the most significant cross-bicorrelation windows (28.1%) and COLCAP that presents the less percentage of windows (14.0%). We also plot the returns and the $(1 - p)$-values of the significant cross-bicorrelation windows. As can be seen in the prices and returns plots, it is clear how for the year 2008 and the year 2011, the SP500 falls in prices and has higher volatility before the other indexes.

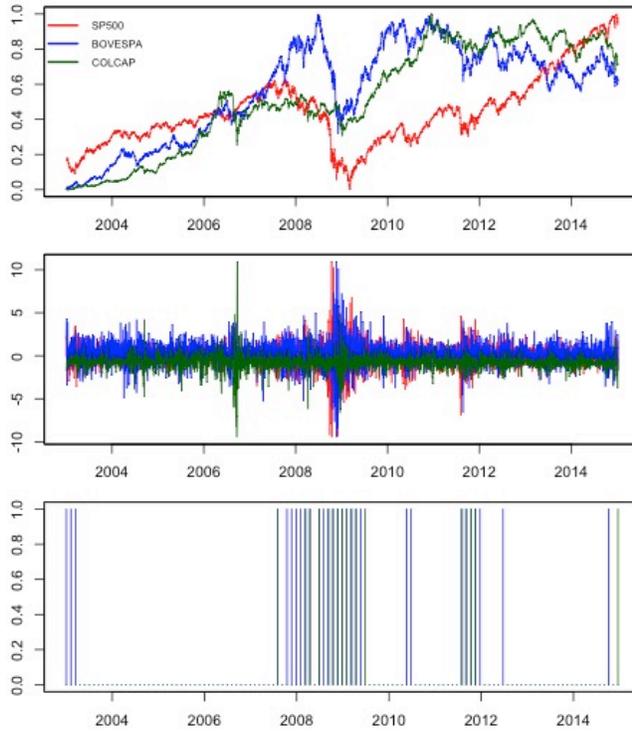

Figure 3. Plot of normalized prices, returns and ($1-p$)-values of the significant cross-bicorrelation windows (top, middle, bottom, respectively) for SP500 (red), BOVESPA (blue) and COLCAP (green).
Source: The authors.

## 5 Concluding remarks

In this paper we have successfully applied the Brooks and Hinich cross-bicorrelation test to uncover the cross-covariances and cross-correlations between USA's Standard and Poor's 500 (SP500), used as benchmark, and six Latin American stock markets indexes: Mexico (BMV), Brazil (BOVESPA), Chile (IPSA), Colombia (COLCAP), Peru (IGBVL) and Argentina (MERVAL). We have found windows of nonlinear dependence between SP500 and the Latin American stock markets, some of which coincide with periods of crisis, giving way to a possible contagion or interdependence interpretation.

This nonlinearity test presents several advantages, since it is able to raise any form of nonlinear dependence of the third-order statistics between two series. It also provides a helpful tool for researchers to determine the functional form of the nonlinear relationship between the series by determining in which direction the cross-bicorrelations flow and which of the lags are significant. Given that this test allows the researcher to determine the third-order nonlinear dependency forms between pairs of series, it can be used as an additional tool to the Granger causality test.

We have identified some moments of cross-bicorrelations that might be interesting to explore from a deeper economical point of view and it is left as work in progress.

## 5 Acknowledgements

The authors would like to acknowledge Jorge Ahumada García (MSc in Economics student, ITAM) for his help providing some of the data used in this paper and Itzel Cano (BA student in Finance, UP) for her help with some of the tables. All errors remain the sole responsibility of the authors.

**Semei Coronado** obtained a Ph.D. degree in Business and Economics from the University of Guadalajara, in Guadalajara, Mexico. He is a Research Professor in the Department of Quantitative Methods at the University of Guadalajara. He is currently a member of the Mexican National System of Researchers (Level I, CONACYT) where his research areas of interest are: Time series, emerging market finance and applied statistics.

**Omar Rojas,** received his PhD in Mathematics in 2009 from La Trobe University, Melbourne, Australia. He is a Research Professor and Research Director at the School of Business and Economics at Universidad Panamericana, Guadalajara, Mexico. His research areas of interest are: nonlinear time series and multivariate statistical methods applied to business and construction. He is a member of Mexican National System of Researchers.
ORCID: orcid.org/0000-0002-0681-3833

**Rafael Romero-Meza** received a PhD in Business Administration from Boston University. His is director of finance of PKF Chile Finanzas Corporativas, Chile. Dr. Romero-Meza has over twenty academic articles in Applied Economics Letters, Macroeconomic Dynamics, among others. He participates in the editorial boards of several journals: Economics Letters, INNOVAR, Emerging Markets, His research and teaching interests are in Emerging markets, Efficient markets, nonlinear time series, corporative finance.

**Francisco Venegas-Martinez** was a post-doctoral researcher in Finance at Oxford University. He received his PhD in Mathematics and a second PhD in economics from Washington State University. He is a professor in Instituto Politécnico Nacional, Mexico. Dr Vengas-Martínez is member of the Mexican National System of Researchers (Level III, CONACYT). He participates in more than 20 editorial boards and scientific research journals, in Mexico and internationally Dr Venegas-Martínez has over hundred academic articles in The Brazilian Journal of Probability, Journal of the Inter-American Statistics and Econometrics, Journal of Economic Dynamics and Control, International Journal of Theoretical and Applied Finance, Journal of Economic Modeling, Journal of Development Economics, among others. His research areas of interest are Stochastic Process, Econometrics, Time Series, and Economic Development.